\documentclass[aps,pra,reprint]{revtex4-2}

\usepackage{graphicx}
\usepackage{latexsym}
\usepackage{dcolumn}
\usepackage{bm}
\usepackage[utf8]{inputenc}
\usepackage[T1]{fontenc}
\usepackage{tabularx}
\usepackage{hyperref}
\usepackage{color}
\usepackage{placeins}
\usepackage{float}
\usepackage{amsmath}
\usepackage{gensymb}

\begin{document}

\title{Atomistic modeling of thermal effects in focused electron beam induced deposition of Me$_2$Au(tfac)}

\author{Alexey Prosvetov}
\email{prosvetov@mbnexplorer.com}
\affiliation{MBN Research Center, Altenh\"oferallee 3, 60438 Frankfurt am Main, Germany}
\author{Alexey V. Verkhovtsev}
\email{verkhovtsev@mbnexplorer.com}
\affiliation{MBN Research Center, Altenh\"oferallee 3, 60438 Frankfurt am Main, Germany}
\author{Gennady Sushko}
\affiliation{MBN Research Center, Altenh\"oferallee 3, 60438 Frankfurt am Main, Germany}
\author{Andrey V. Solov'yov}
\affiliation{MBN Research Center, Altenh\"oferallee 3, 60438 Frankfurt am Main, Germany}

\date{\today}

\begin{abstract}
The role of thermal effects in the focused electron beam induced deposition (FEBID) process of Me$_2$Au(tfac) is studied by means of irradiation-driven molecular dynamics simulations. The FEBID of Me$_2$Au(tfac), a commonly used precursor molecule for the fabrication of gold nanostructures, is simulated at different temperatures in the range of $300-450$~K. The deposit's structure, morphology, growth rate, and elemental composition at different temperatures are analyzed. The fragmentation cross section for Me$_2$Au(tfac) is evaluated on the basis of the cross sections for structurally similar molecules. Different fragmentation channels involving the dissociative ionization (DI) and dissociative electron attachment (DEA) mechanisms are considered. The conducted simulations of FEBID confirm experimental observations that deposits consist of small gold clusters embedded into a carbon-rich organic matrix. The simulation results indicate that accounting for both DEA- and DI-induced fragmentation of all the covalent bonds in Me$_2$Au(tfac) and increasing the amount of energy transferred to the system upon fragmentation increase the concentration of gold in the deposit. The simulations predict an increase in Au:C ratio in the deposit from 0.18 to 0.25 upon the temperature increase from 300~K to 450~K, being within the range of experimentally reported values.
\end{abstract}

\maketitle

\widowpenalty=10000
\section{Introduction}
\label{Intro}

Focused Electron Beam Induced Deposition (FEBID) is a technology for the controllable fabrication of complex nanostructures with nanometer resolution \cite{Utke_book_2012,DeTeresa-book2020,Winkler_2019_JAP_review, Huth_2021_JAP_review}. The FEBID process consists of the deposition of organometallic precursor molecules on a substrate and irradiation of the adsorbed molecules by a focused keV-energy electron beam. Electron-induced decomposition releases organic ligands resulting in clusterization of the precursor's metallic component on a surface. The lateral size of the resulting deposit is comparable to that of the incident electron beam (typically, $\sim$1--10 nanometers) \cite{Plank2020}.

The FEBID process involves a complex interplay of different phenomena taking place on different temporal and spatial scales: (i) adsorption, diffusion and desorption of precursor molecules on/from a substrate; (ii) transport of primary, secondary and backscattered electrons; (iii) electron-induced dissociation of the adsorbed precursor molecules; and (iv) follow-up chemical transformations.

The atomistic modeling of the FEBID process has become possible recently by means of Irradiation-Driven Molecular Dynamics (IDMD) \cite{Sushko2016}, a novel and general methodology for computer simulations of irradiation-driven transformations of complex molecular systems. This method enables the atomistic-level description of nanostructures grown by FEBID \cite{Sushko2016,MBNbook_Springer_2017, DeVera2020} with accounting for chemical transformations of adsorbed molecular systems \cite{Sushko2016a} irradiated with a focused electron beam.

Within the IDMD framework, various quantum processes occurring in an irradiated system (e.g. covalent bond breakage induced by ionization or electron attachment) are treated as random, fast and local transformations incorporated into the classical MD framework in a stochastic manner with the probabilities elaborated on the basis of quantum mechanics \cite{Sushko2016}.
Major transformations of irradiated molecular systems (such as molecular topology changes, redistribution of atomic partial charges, or alteration of interatomic interactions) are simulated by means of MD with the reactive CHARMM (rCHARMM) force field \cite{Sushko2016a} using the advanced software packages MBN Explorer \cite{Solovyov2012} and MBN Studio \cite{Sushko2019}. MBN Explorer is a multi-purpose software package for multiscale simulations of the structure and dynamics of complex Meso-Bio-Nano (MBN) systems \cite{MBNbook_Springer_2017}. MBN Studio is a powerful multi-task toolkit used to set up and start MBN Explorer calculations, monitor their progress, examine calculation results, visualize inputs and outputs, and analyze specific characteristics determined by the output of simulations \cite{Sushko2019}.

A detailed overview of the computational workflow for IDMD-based simulations of the FEBID process has been presented in the recent study \cite{Prosvetov2021}, and the methodology was utilized to simulate the FEBID of Pt(PF$_3$)$_4$ precursor molecules. The simulations carried out in Ref.~\cite{Prosvetov2021} described the initial stage of nanostructure growth, including nucleation of metal atoms, formation of small metal clusters on a surface, their aggregation and, eventually, the formation of a dendritic metal nanostructure.
In the follow-up study \cite{Prosvetov2021a} with Fe(CO)$_5$ precursor molecules, the variation of the deposit's structure, morphology and metal content at different irradiation and replenishment conditions of the FEBID process was investigated. It was demonstrated that either a nanogranular deposit consisting of small-size metal clusters surrounded by organic ligands or a single dendrite-like structure with the size corresponding to the primary electron beam is formed depending on the beam current. The aforementioned studies \cite{Sushko2016, DeVera2020, Prosvetov2021, Prosvetov2021a} have demonstrated the successful application of the IDMD methodology for the atomistic simulations of FEBID.

Investigation of the phenomena that govern the formation and growth of nanostructures in the FEBID process is a complex multi-parameter problem. Indeed, different precursor molecules, substrate types as well as irradiation, replenishment and post-processing conditions
can be explored to fabricate deposits with the optimal geometries and compositions.
However, due to the complexity of the problem, not all the mentioned aspects of the FEBID process have been explored so far by means of IDMD simulations.
One of the parameters influencing the properties of FEBID-grown deposits is the operational temperature of the FEBID process \cite{Mulders2011,Rosenberg2012,DeTeresa2019a,Huth2020}.
The temperature effects arising during the FEBID process have been considered by means of the continuum diffusion-reaction model \cite{Toth2015}. However, this approach cannot provide atomistic details of the deposit's structure. Up to now, the thermal effects during FEBID have not been studied by means of IDMD.

\begin{figure}[t]
\includegraphics[width=0.4\textwidth]{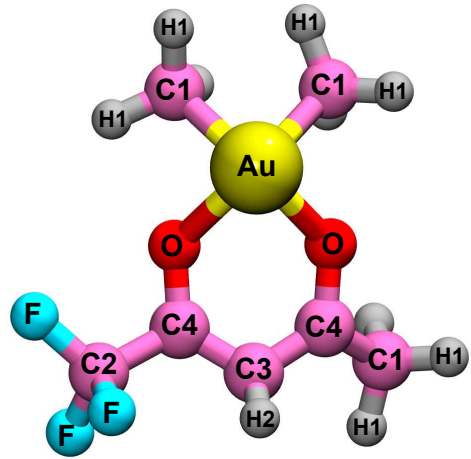}
\caption{Optimized geometry of a Me$_2$Au(tfac) molecule considered in this study. The optimization calculation has been performed by means of MBN Explorer using the interatomic potentials given by Eqs.~(\ref{Eq. Morse})--(\ref{Eq. Lennard-Jones}). Different atom types are indicated. The corresponding bonded and angular interactions are listed in Table~\ref{Table:CovBonds}. }
\label{Fig:molecule}
\end{figure}

\begin{sloppypar}
Different types of precursor molecules have been proposed for FEBID applications, see reviews~\cite{Utke2008, Barth2020_JMaterChemC, Utke2022} and references therein. For example, one can mention metal carbonyls (e.g. Fe(CO)$_5$, W(CO)$_6$ or Co$_2$(CO)$_8$), phosphines (e.g. Pt(PF$_3$)$_4$), halides (e.g. Pt(NH$_3$)$_2$Cl$_2$ or Pt(CO)$_2$Cl$_2$), cyclopentadienyl complexes (e.g. MeCpPtMe$_3$) and $\beta$-diketonates (e.g. Cu(hfac)$_2$).
Some precursors, such as metal carbonyls considered in the previous IDMD-based studies \cite{Sushko2016, DeVera2020, Prosvetov2021a}, have relatively simple geometries where one or two metal atoms are linked to small ligands of a same type. Other precursors, such as $\beta$-diketonates (e.g. dimethyl-gold-trifluoroacetylacetonate, Me$_2$Au(tfac), shown in Fig.~\ref{Fig:molecule}), have more complex geometries with many different atom types and different covalent interactions, opening a broad spectrum of electron irradiation induced fragmentation channels. At the same time, the available data on the absolute fragmentation cross sections for such complex precursors are very limited or do not exist. A detailed comparative study on dissociative electron attachment (DEA) to the isolated diketones (acetylacetone -- \textit{acac}, trifluoroacetylacetone -- \textit{tfac}, and hexafluoroacetylacetone -- \textit{hfac}) was presented in Ref.~\cite{Omarsson2014}. A comparison of the experimentally measured electron-induced fragmentation of acetone and acac was performed in Ref.~\cite{Warneke2015a}. Decomposition of the metal-acac complexes with Cu, Mn and Zn atoms irradiated with low-energy ($0-10$~eV) electrons was studied in Refs.~\cite{Kopyra2018,Kopyra2020,Kopyra2020a}. However, only relative yields of fragments created due to the electron-induced fragmentation of parent molecules were reported in the cited studies. Electron-induced surface reactions and products, reaction kinetics and structure of FEBID-grown deposits for Me$_2$Au(acac) precursors adsorbed onto solid substrates were discussed in Ref.~\cite{Wnuk2010}.
\end{sloppypar}

\begin{sloppypar}
$\beta$-diketonates are an important class of organometallic precursors in FEBID. In particular, $\beta$-diketonate complexes with Au atoms (Me$_2$Au(acac) = C$_7$H$_{13}$O$_2$Au, Me$_2$Au(tfac) = C$_7$H$_{10}$F$_3$O$_2$Au, and Me$_2$Au(hfac) = C$_7$H$_{7}$F$_6$O$_2$Au) are among the main FEBID precursors used for the fabrication of gold nanostructures \cite{Graells2007, Kuhness2021}. A $\beta$-diketone ligand encloses the metal atom forming a rigid 6-membered ring (see Fig.~\ref{Fig:molecule}), which efficiently protects the metal atom against chemical reactions. As a result, as-grown Au deposits produced using these precursors under normal conditions usually contain only $\sim 5-20$~at.\% of metal and are contaminated with a high percentage of carbon atoms \cite{Barth2020_JMaterChemC,Utke2022}.
\end{sloppypar}

Different purification techniques have been developed to increase the Au content \cite{Botman2009a}, for example post-growth annealing \cite{DosSantos2018}, deposition at elevated temperatures \cite{Mulders2011}, FEBID under reactive atmosphere, e.g. water vapor or O$_2$, alongside the deposition or the subsequent annealing \cite{Shawrav2016a, Mansilla2016}. Overall, a higher metal content can be achieved by promoting the release of ligands and adjusting the environment to enable the formation of volatile chemical products. Nevertheless, the efficiency of the purification methods and, particularly, thermal treatment of FEBID-grown deposits varies greatly for different precursors. Moreover, results for the same precursor differ in the studies reported by different research groups \cite{Botman2009a, Barth2020_JMaterChemC, Utke2022}.

In Ref.~\cite{Koops1996} the gold-containing nanostructure growth during FEBID of Me$_2$Au(tfac) was studied experimentally. An increase of Au content in the deposited material from $1-15$~at.\% up to 24~at.\% during the substrate heating up to 373~K was reported.
Mulders \textit{et al.} \cite{Mulders2011} observed an increase in Au content from ca. 20 to 30~at.\% with increasing the temperature from 300 to 450~K during FEBID of Me$_2$Au(acac).
At the same time, other experimental studies indicated that the effect of post-deposition thermal processing of FEBID-grown deposits of Me$_2$Au(acac) is strongly influenced by the accompanying gas. Botman \textit{et al.} \cite{Botman2006} reported that the Au content of $\sim$8~at.\% in the deposit did not increase upon annealing at different elevated temperatures up to 673~K in an N$_2$ atmosphere and annealing at 473~K in air. However, when the annealing was performed in an O$_2$ atmosphere, the average Au concentration increased gradually up to $\sim$13~at.\% at 523~K and raised up to 60~at.\% at 673~K \cite{Botman2006}. Post-growth annealing of the structures obtained after FEBID of Me$_2$Au(acac) and Cu(hfac)$_2$ precursors at 573~K did not increase the initial metal content of $\sim$5~at.\% \cite{DosSantos2018}.

In the present study, the role of thermal effects during the FEBID process is investigated at the atomistic level by means of IDMD simulations.
The Me$_2$Au(tfac) precursor molecule is considered as an illustrative case study. The FEBID of Me$_2$Au(tfac) is simulated at different temperatures in the range of $300-450$~K, and the deposit's structure, morphology, growth rate, and elemental composition at different temperatures are analyzed. The simulations show that the deposit consists of small metal clusters containing several gold atoms embedded into an organic matrix. An increase in Au:C ratio in the deposits from $\sim$0.18 to $\sim$0.25 is observed when the temperature increases from 300 to 450~K, which is within the range of experimentally reported data.

The absolute cross section of electron-impact induced fragmentation of Me$_2$Au(tfac) is evaluated by different methods. Four different approximations for the fragmentation cross section are considered and compared. In the simplest approximation, the total cross section accounts only for the dissociative ionization (DI)-induced cleavage of covalent bonds between the gold atom and the ligands. The most complete approximation for the fragmentation cross section accounts for the contribution of DI and DEA processes in the cleavage of covalent bonds between the gold atom and the ligands, as well as for the bond cleavage within the ligands. The yields of created atomic and molecular fragments are compared for the considered approximations for the fragmentation cross section and for different values of the energy deposited into the system upon a covalent bond breakage, $E_d$. The simulation results and their analysis indicate that accounting for both DEA- and DI-induced fragmentation of all the covalent bonds in Me$_2$Au(tfac) and increasing $E_d$ result in growing the concentration of metal content in the deposits. The simulated concentration of gold in the deposit and the dependence of the deposit's growth rate on temperature are within the range of experimental values reported for Me$_2$Au(tfac) and structurally similar precursor molecules.


\section{Computational methodology}
\label{Methods}

Computer simulations of the FEBID process of Me$_2$Au(tfac) on a fully hydroxylated silica (SiO$_2$-H) substrate have been performed by means of the MBN Explorer software package \cite{Solovyov2012}. The MBN Studio toolkit \cite{Sushko2019} has been utilized to create the systems, prepare all necessary input files and analyze simulation outputs. The simulations have followed the multi-step computational protocol described in Ref.~\cite{Prosvetov2021}.

\subsection{Interatomic interactions}

Interatomic interactions involving the precursor molecules and their molecular fragments have been described by means of the reactive CHARMM (rCHARMM) force field \cite{Sushko2016a}. rCHARMM permits simulations of systems with dynamically changing molecular topologies, which is essential for modeling the precursor fragmentation \cite{DeVera2019} and the formation of metal-containing nanostructures \cite{Sushko2016, DeVera2020, Prosvetov2021, Prosvetov2021a}. A detailed description of rCHARMM is given in Ref.~\cite{Sushko2016a}, see also a recent review \cite{Verkhovtsev2021}.

The radial part of bonded interactions is described in rCHARMM by means of the Morse potential:
\begin{equation}
U^{{\rm bond}}(r_{ij}) = D_{ij} \left[ e^{-2\beta_{ij}(r_{ij} - r_0)} - 2e^{-\beta_{ij}(r_{ij} - r_0)} \right] \ .
\label{Eq. Morse}
\end{equation}
Here $D_{ij}$ is the dissociation energy of the bond between atoms $i$ and $j$, $r_0$ is the equilibrium bond length, and $\beta_{ij} = \sqrt{k_{ij}^{r} / D_{ij}}$ (with $k_{ij}^{r}$ being the bond force constant) determines the steepness of the potential. The bonded interactions are truncated at a user-defined cutoff distance that characterizes the distance beyond which the bond becomes broken and the molecular topology of the system changes. 

The rupture of covalent bonds in the course of simulation employs the reactive potential for valence angles:
\begin{equation}
    U^{{\rm angle}}(\theta_{ijk}) = 2 k^\theta_{ijk} \, \sigma(r_{ij}) \, \sigma(r_{jk}) \left[ 1 - \cos(\theta_{ijk}-\theta_0 )  \right] \ ,
\label{Eq. Angles}
\end{equation}
where $\theta_0$ is the equilibrium angle, $k^{\theta}$ is the angle force constant, and the function $\sigma(r_{ij})$ describes the effect of bond breakage \cite{Sushko2016a}:
\begin{equation}
    \sigma(r_{ij}) = \frac{1}{2} \left[1-\tanh(\beta_{ij}(r_{ij}-r_{ij}^*))  \right] \ .
\label{Eq. Rupture_param}
\end{equation}
Here $r_{ij}^*=(R^{{\rm vdW}}_{ij}+r_0)/2$, with $r_0$ being the equilibrium distance between two atoms involved in the angular interaction and $R^{{\rm vdW}}_{ij}$ being the van der Waals radius for those atoms.

\begin{table*}[t!]
\centering
\caption{Parameters of the covalent bonded and angular interactions, Eqs.~(\ref{Eq. Morse}) and (\ref{Eq. Angles}), for a Me$_2$Au(tfac) molecule, used in the simulations. The corresponding atom types are shown in Fig.~\ref{Fig:molecule}.}
\begin{tabular}{p{3.5cm}p{2cm}p{4cm}p{3.5cm}}
\hline
Bond     &   $r_0$~(\AA)   &  $D_{ij}$~(kcal/mol)  &  $k_{ij}^{r}$~(kcal/mol \AA$^{-2}$)  \\
\hline
F--C2   &    1.43    &    154.3   &  371.7    \\
C4--C2  &    1.57    &    144.5   &  538.5    \\
C4--O   &    1.34    &    215.9   &  538.5    \\
C4--C3  &    1.39    &    222.0   &  683.0    \\
C3--H2  &    1.16    &    171.2   &  387.0    \\
C1--H1  &    1.16    &    171.2   &  387.0    \\
Au--O   &    2.16    &     45.0   &  133.4    \\
Au--C1  &    2.06    &     81.2   &  206.4    \\
\hline
Angle  &  $\theta_0$~(deg.)  & $k_{ijk}^{\rm {\theta}}$~(kcal/mol rad$^{-2}$)   \\
\hline
C3--C4--C1   &  118.2   &  48.0   \\
C3--C4--C2   &  118.2   &  56.0   \\
C3--C4--O    &  128.0   &  56.0   \\
C1--C4--O    &  113.0   &  75.0   \\
C2--C4--O    &  113.0   &  75.0   \\
C4--C3--C4   &  120.0   &  65.0   \\
C4--C2--F    &  112.0   &  50.0   \\
C4--C1--H1   &  110.0   &  42.0   \\
C1--Au--C1   &   88.3   &  34.0   \\
F--C2--F     &  107.0   &  50.0   \\
C4--C3--H2   &  117.0   &  42.0   \\
C4--O--Au    &  125.0   &  42.0   \\
O--Au--O     &   86.5   &  42.0   \\
H1--C1--H1   &  109.0   &  42.0   \\
Au--C1--H1   &  107.0   &  42.0   \\
C1--Au--O    &   92.5   &  42.0   \\
\hline
\end{tabular}
\label{Table:CovBonds}
\end{table*}

\begin{table}[ht!]
\caption{Parameters of the Lennard-Jones potential, Eq.~(\ref{Eq. Lennard-Jones}), describing the van der Waals interaction between atoms of a Me$_2$Au(tfac) precursor molecule, its fragments and atoms of the substrate.
}
\centering
\begin{tabular}{c|c|c|c}
	Atom type	& $\varepsilon_i$ (kcal/mol) & $r_i^{{\rm min}}/2$~(\AA)  & Ref. \\
\hline
Au             & 5.29  &  1.48  &  \cite{pohjolainen2016unified}  \\
F              & 0.07  &  1.47  &  \cite{SwissParam_paper} \\
C1--C4         & 0.06  &  2.02  &  \cite{SwissParam_paper} \\
O              & 0.10  &  1.65  &  \cite{SwissParam_paper} \\
H1, H2         & 0.04  &  1.34  & \cite{SwissParam_paper} \\
Si             & 0.31  &  2.14  &  \cite{Mayo1990} \\
O$_{\rm sub}$  & 0.10  &  1.70  &  \cite{Mayo1990} \\
H$_{\rm sub}$  & 0.08  &  0.30  &  \cite{Mayo1990} \\
\hline
\end{tabular}
\label{Table:van_der_Waals} 	
\end{table}

The initial geometry of a Me$_2$Au(tfac) molecule has been determined via density-functional theory (DFT) calculations using the Gaussian software package \cite{Gaussian09} and then optimized using MBN Explorer. The optimized geometry of Me$_2$Au(tfac) is shown in Fig.~\ref{Fig:molecule}.
The rCHARMM parameters for Me$_2$Au(tfac) have been determined from a series of DFT-based potential energy surface scans, following the procedure described in the earlier studies~\cite{DeVera2019, Prosvetov2021}.
The parameters of the bonded and angular interactions for Me$_2$Au(tfac) are listed in Table~\ref{Table:CovBonds}.

In the present simulations, we consider the physisorption of precursor molecules on a SiO$_2$-H substrate. Therefore, the molecules do not form covalent bonds with atoms of the substrate but interact with them via van der Waals forces described by means of the Lennard-Jones potential:
\begin{equation}
    U_{{\rm LJ}}(r_{ij})=\varepsilon_{ij} \, \left [ \left (\frac{r^{{\rm min}}}{r_{ij}} \right )^{12}-2\left (\frac{r^{{\rm min}}}{r_{ij}}\right )^6 \right ]  ,
    \label{Eq. Lennard-Jones}
\end{equation}
where $\varepsilon_{ij}=\sqrt{\varepsilon_i \, \varepsilon_j}$ and $r^{{\rm min}} = (r^{{\rm min}}_i+r^{{\rm min}}_j)/2$.
Parameters of the Lennard-Jones potential for gold atoms
have been taken from Ref.~\cite{pohjolainen2016unified}. Parameters for other atoms of the precursor molecule have been generated using the SwissParam web-service \cite{SwissParam_paper}. Parameters for atoms of the substrate have been taken from Ref.~\cite{Mayo1990}. All these parameters are summarized in Table~\ref{Table:van_der_Waals}.
The bonded interaction between Au atoms in the formed deposits has been described by means of the many-body Gupta potential \cite{Gupta_1983_PRB.23.6265} with the parameters taken from Ref.~\cite{Cleri1993}.
Following the earlier IDMD-based studies of FEBID  \cite{Sushko2016,DeVera2020, Prosvetov2021}, the substrate has been considered frozen to speed up the simulations.

\subsection{The system formation}

In this study, a layer of Me$_2$Au(tfac) molecules with the size $20~\textrm{nm} \times 20~\textrm{nm}$ has been created by means of MBN Studio \cite{Sushko2019}, optimized, deposited on the SiO$_2$-H substrate and equilibrated at different temperatures (300, 350, 400 and 450~K) for 0.5~ns using the Langevin thermostat with a damping time of 0.2~ps. The number of precursor molecules added at each temperature considered has been determined through the equilibrium surface density of Me$_2$Au(tfac) evaluated via the adsorption-desorption rates from the continuum model of FEBID \cite{Toth2015}. The surface densities of Me$_2$Au(tfac) have been calculated as follows.

According to the kinetic theory of gases, the uniform molecular flux $F_{\rm p}$ impinging on a surface placed in a chamber with pressure $P_{\rm p}$ is given by \cite{LANDAU1980111}:
\begin{equation}
F_{\rm p} = \frac{P_{\rm p}}{\sqrt{2\pi \, m_{\rm p} \, k_{\rm B} T_{\rm p}}} \ ,
\label{Eq. GasFlux}
\end{equation}
where $T_{\rm p}$ is the gas temperature, $m_{\rm p}$ is the mass of the precursor molecule, and $k_{\rm B}$ is the Boltzmann constant.
The rate of newly physisorbed precursor molecules $\varphi_{\rm p}$ is defined as:
\begin{equation}
\varphi_{\rm p} = s_{\rm p} F_{\rm p} (1 - A_{\rm p} n_{\rm p}) \ ,
\label{Eq. Adsorption}
\end{equation}
where $s_{\rm p}$ is the precursor sticking coefficient, commonly set equal to 1, $d_{\rm p}$ is the diameter of a circle circumscribing the molecule, $A_{\rm p} = \pi d_{\rm p}^2/4$ is the surface area covered by one precursor molecule, and $n_{\rm p}$ is the surface density of precursors.
The thermal desorption rate $k_{\rm p}$ is calculated according to:
\begin{equation}
k_{\rm p} = \kappa_{\rm p} \exp{\left(-\frac{E_{\rm p}}{k_{\rm B} T}\right)} \ ,
\label{Eq. Desorption}
\end{equation}
where $\kappa_{\rm p}$ is the desorption attempt frequency, $E_{\rm p}$ is the desorption energy, and $T$ is the substrate temperature.

The surface density of precursor molecules is given by the equation
\begin{equation}
\frac{{\rm d}n_{\rm p}}{{\rm d}t} = \varphi_{\rm p} - n_{\rm p} k_{\rm p} \ ,
\label{Eq. SurfaceConcentration}
\end{equation}
with the solution for $n_{\rm p}(t)$:
\begin{equation}
\label{Eq. SurfaceConcentrationSolution}
    n_{\rm p}(t) = \frac{(1-e^{-(s_{\rm p} F_{\rm p} A_{\rm p} + k_{\rm p}) t }) s_{\rm p} F_{\rm p}  }{s_{\rm p} F_{\rm p} A_{\rm p} + k_{\rm p}}.
\end{equation}
The solution of Eq.~(\ref{Eq. SurfaceConcentration}) at $t \gg \left( s_{\rm p} F_{\rm p} A_{\rm p} + k_{\rm p} \right)^{-1}$ (which is equal to $\sim$0.1--10~$\mu$s for the studied temperature range $T = 300-450$~K) gives the steady-state surface density of precursors:
\begin{equation}
n_{\rm p_0} \approx \frac{s_{\rm p} F_{\rm p}}{s_{\rm p} F_{\rm p} A_{\rm p} + k_{\rm p}} \ .
\label{Eq. SteadyStateConcentration}
\end{equation}
The values of $n_{\rm p_0}$ calculated using Eq.~(\ref{Eq. SteadyStateConcentration}) for the surface temperatures $T = 300$, 350, 400 and 450~K are equal to $n_{\rm p_0} = 2.6$, 2.3, 0.8 and 0.1 molecules/nm$^2$, respectively.
The calculations have been performed with the values of gas temperature $T_{\rm p} = 300$~K and pressure $P_{\rm p} = 20$~Pa.
The mass of a Me$_2$Au(tfac) molecule is $m_{\rm p} = 380$~a.m.u. and the diameter $d_{\rm p} \approx 0.7$~nm. The values of desorption attempt frequency  $\kappa_{\rm p} = 1.0 \times 10^{-14}$~s$^{-1}$ and desorptioin energy $E_{\rm p} = 0.7$~eV have been approximated by the $\kappa_{\rm p}$ and $E_{\rm p}$ values for organic molecules of a similar size \cite{Fichthorn2002,Cullen2015a}.

The chosen value of $P_{\rm p}$ is several times larger than the vapor pressure of Me$_2$Au(tfac) at room temperature, equal to 7~Pa \cite{Ohta2001}, and $1-2$ orders of magnitude larger than the values of precursor gas pressure considered in Ref.~\cite{Cullen2015a} within the continuum diffusion-reaction model.
The steady-state precursor surface density $n_{\rm p_0}$ calculated using Eq.~(\ref{Eq. SteadyStateConcentration}) for $T = 300$~K decreases only by 10\% when the pressure $P_{\rm p}$ decreases from 20~Pa to 1~Pa. At higher temperatures, the thermal desorption (the second term in the denominator of Eq.~(\ref{Eq. SteadyStateConcentration})) becomes dominant, leading to a faster decrease of the equilibrium surface density for smaller $P_{\rm p}$ values. The resulting system is very sparse, with an average distance between the deposited molecules on the order of several nanometers and a surface coverage close to zero. Considering smaller values of $P_{\rm p}$ and $n_{\rm p_0}$ at elevated temperatures would require running very long MD simulations (on the $\mu$s time scale), which is a challenging computational task.
Therefore, in the present study we have considered a higher $P_{\rm p}$ value that enables us to simulate the deposit's growth at elevated temperatures on the computationally feasible timescale of hundreds of nanoseconds.

IDMD simulations of the FEBID process use the information on the space resolved fragmentation probability per unit time, $P(x,y)$. The probability is calculated using the spatial distribution of flux density of primary (PE), secondary (SE) and backscattered (BSE) electrons \cite{DeVera2020} and the absolute fragmentation cross section of the precursor molecules:
\begin{eqnarray}
    P(x,y) &=& \sigma_{\rm frag}(E_0) J_{\rm PE}(x,y,E_0) \\ 
    &+& \sum_i \sigma_{\rm frag}(E_i) [J_{\rm SE}(x,y,E_i) + J_{\rm BSE}(x,y,E_i) ] \nonumber \ .
\label{Eq. Frag_Probability_total}
\end{eqnarray}
Here $\sigma_{\rm frag}(E)$ is the energy-dependent precursor fragmentation cross section, $E_i < E_0$ is the electron energy discretized in steps of 1~eV, and $J_{\rm PE/SE/BSE}(x,y,E_i)$ are the flux densities of PE, SE and BSE with energy $E_i$ at the point ($x$,$y$), respectively.
The spatial distribution of the electron flux density employed in the calculation of $P(x,y)$ was obtained previously \cite{DeVera2020} using the track-structure Monte Carlo code SEED for a cylindrical PE beam with a radius of 5~nm and energy $E_0 =  1$~keV.

\subsection{Fragmentation cross section}
\label{sec:fragm_CS}

The main mechanisms of molecular fragmentation are dissociative electron attachment (DEA) at low electron energies below the ionization potential of the molecule (typically below $\sim$10~eV) and dissociative ionization (DI) at higher electron energies.
To the best of our knowledge, there is no data in the literature on the absolute fragmentation cross section of Me$_2$Au(tfac). Therefore, the fragmentation cross section $\sigma_{\rm frag}(E)$ has been evaluated based on the compilation of data available for the fragmentation cross sections of structurally similar molecules and smaller functional groups of Me$_2$Au(tfac).

\begin{figure*}[t!]
\includegraphics[width=0.75\textwidth]{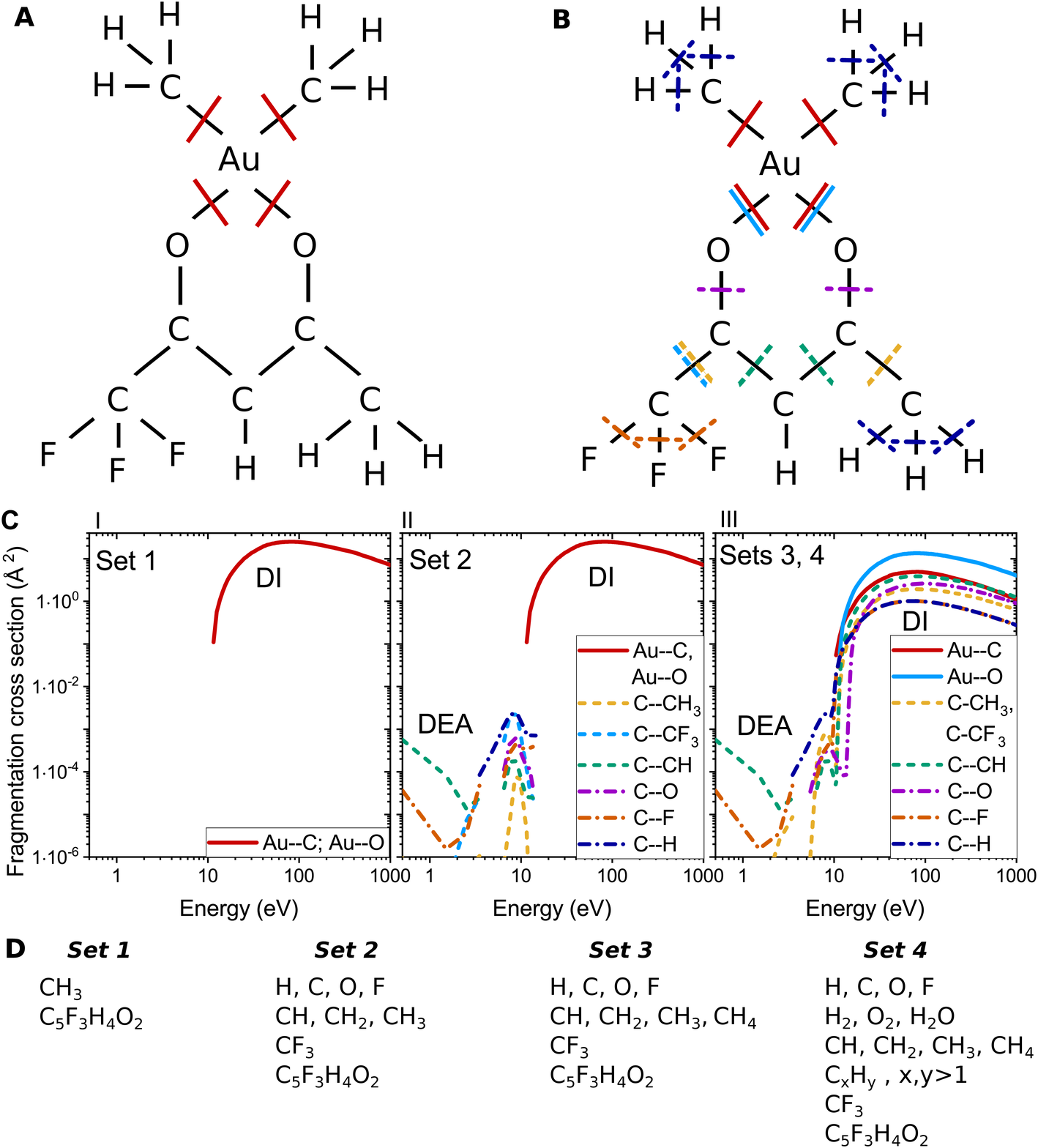}%
\caption{Panels \textbf{A} and \textbf{B} show a schematic representation of the Me$_2$Au(tfac) molecule. Colored lines indicate the covalent bonds whose electron-induced dissociation is considered in the simulations using Set~1 (panel~A) and Sets 2, 3 and 4 (panel~B) of the fragmentation parameters (see the main text for details). Panel~\textbf{C} shows four sets of the electron impact fragmentation cross section of Me$_2$Au(tfac) used in the simulations. \textbf{C-I:} Set 1 accounts for the fragmentation of metal-ligand bonds due to DI, see Eqs.~(\ref{Eq. TotalIonizationCS}) and (\ref{Eq. tfac_IonizationCS}). \textbf{C-II:}  Set 2 accounts for bond breakage within the CH$_3$ and \textit{tfac} ligands as a result of DEA and breakage of the metal-ligands bonds due to DI. \textbf{C-III:} Sets 3 and 4 take into account bond breakage due to DEA and DI for all bonds in the Me$_2$Au(tfac) molecule, see Eqs.~(\ref{Eq. TotalIonizationCS})--(\ref{Eq. PartialDIAu}). Additionally, chemical reactions involving produced atomic and molecular fragments have been accounted for in simulations performed with Set~4. Panel~\textbf{D} indicates atomic and molecular fragments produced in the simulations using the considered sets of fragmentation parameters.}
\label{Fig:Cross-section}
\end{figure*}

The electron-impact induced fragmentation of a Me$_2$Au(tfac) molecule is governed by the contributions from various dissociation channels. As such, determination of the fragmentation cross section for Me$_2$Au(tfac) is a non-trivial task. In the present study, the cross section $\sigma_{\rm frag}(E)$ has been evaluated by different methods. Four different approximations for the total fragmentation cross section (denoted hereafter as ``sets'') accounting for various dissociation channels have been considered for comparison. The first set is based on the simplest approximation accounting for DI-induced cleavage of the bonds between the gold atom and the ligands. The most detailed approximation considered in this study accounts for DEA and DI-induced cleavage of all the bonds in Me$_2$Au(tfac) with follow-up chemistry involving the produced fragments. The summary of the considered sets is presented in Fig.~\ref{Fig:Cross-section}.

Experimental studies of electron-impact ionization and fragmentation of organometallic precursor molecules showed that the partial cross section of ionization without fragmentation is significantly (by 1--2 orders of magnitude) smaller than the sum of partial ionization cross sections leading to the emission of ionic fragments \cite{Wnorowski2012, Engmann2013, Thorman2015}.
Therefore, the total ionization cross section can be used as a reasonable approximation for the DI cross section. In the present study, the DI cross section of Me$_2$Au(tfac) has been calculated according to the additivity rule principle \cite{DEUTSCH1994} as a sum of ionization cross sections of the largest functional groups of the molecule.
The following conclusions were made previously from the analysis of the ionization cross sections for two groups of organic molecules -- aldehydes and ketones \cite{Gupta2014, Bull2012a}. First, the shape of ionization cross sections as functions of the projectile electron energy is similar for different molecules of the same group. Second, the maximum value of the ionization cross section is proportional to the number of electrons in a target molecule. The additivity rule-based approach worked well for the studied organic molecules and provided an agreement with experimental data within the range of experimental uncertainties \cite{Gupta2014,  Bull2012a}.

The cross section of DEA has been evaluated in this study as follows. We have utilized the experimental data \cite{Omarsson2014,Warneke2015a} on electron-impact fragmentation mass spectra of \textit{tfac} and structurally similar molecules \textit{acac} and acetone (see Fig.~\ref{Fig:StructuralFormulas}). The absolute DEA cross section for \textit{tfac} has been evaluated by rescaling the spectra from Refs.~\cite{Omarsson2014,Warneke2015a} using the ratio of peak intensities for common molecular fragments. The reported absolute DEA cross section of acetone \cite{Prabhudesai2014} has been used as a reference. The detailed procedure for obtaining the absolute DEA-induced fragmentation cross sections for different bonds of Me$_2$Au(tfac) is described in Sect.~\ref{sec:Fragm_CS_Set2}.

IDMD-based simulations of the FEBID process require the specification of (i) the fragmentation rates for different covalent bonds in the precursor molecule and (ii) the amount of energy $E_d$ deposited into the system during the fragmentation process. This energy is deposited locally into a specific covalent bond of the target and converted into kinetic energy of the two atoms forming the bond \cite{DeVera2019, Sushko2016}.
The choice of $E_d$ may influence the rate of precursor molecule fragmentation \cite{DeVera2020}. For each particular case study the amount of energy transferred by the incident radiation to the system can be evaluated from quantum mechanical calculations of the processes of energy deposition and excitation.
This task, however, goes beyond the scope of this work. Therefore, $E_d$ is considered here as a variable parameter that can be determined from the comparison with experimentally measurable characteristics of the FEBID process, being within the physically justifiable range of values $5~{\rm eV} \lesssim E_d \lesssim 25~{\rm eV}$.
In this study, the value of $E_d$ is varied within the range $400-500$~kcal/mol (from $\sim$17.3 to 21.7~eV) to study the influence of $E_d$ on the Me$_2$Au(tfac) fragmentation process.

As discussed above, the partial cross section of DI, $\sigma_{\rm DI}(E)$, leading to the breakage of specific bonds in the molecule, has been approximated using the total ionization cross section of the molecule, $\sigma_{\rm ion}^{\rm total}(E)$. The latter can be presented as a sum:
\begin{equation}
\sigma_{\rm ion}^{\rm total} (E) = \sigma_{\rm ion}^{\rm fr}(E) + \sigma_{\rm ion}^{\rm nonfr}(E) =  \left[1 + \alpha (E)\right] \sigma_{\rm ion}^{\rm fr}(E) \ ,
\label{Eq. Ionization_CS}
\end{equation}
where $\sigma_{\rm ion}^{\rm nonfr}(E)$ is the cross section of non-dissociative ionization (i.e. ionization without fragmentation), $\sigma_{\rm ion}^{\rm fr}(E)$ is the cross section of ionization with subsequent dissociation, i.e. the DI cross section, and the coefficient $\alpha(E) = \sigma_{\rm ion}^{\rm nonfr}(E)/\sigma_{\rm ion}^{\rm fr}(E)$. Hence, the DI cross section can be written as:
\begin{equation}
\sigma_{\rm ion}^{\rm fr} (E) = \frac{1}{\alpha (E) + 1} \, \sigma_{\rm ion}^{\rm total}(E) .
\label{Eq. DI_CS}
\end{equation}
In the case of a weak covalent bonding, the partial cross section of ionization without fragmentation is much smaller than the DI cross section, $\alpha \ll 1$; hence
\begin{equation}
\label{Eq. DI_CS_approx}
    \sigma_{\rm DI}(E) \approx \sigma_{\rm ion}^{\rm total}(E).
\end{equation}
This approximation has been used in this study to evaluate the partial cross sections of DI, resulting in the cleavage of different covalent bonds in a Me$_2$Au(tfac) molecule, see Sections~\ref{sec:Fragm_CS_Set1} and \ref{sec:Fragm_CS_Set3} below.

\subsubsection{Set 1}
\label{sec:Fragm_CS_Set1}

\begin{figure*}[t!]
\includegraphics[width=0.85\textwidth]{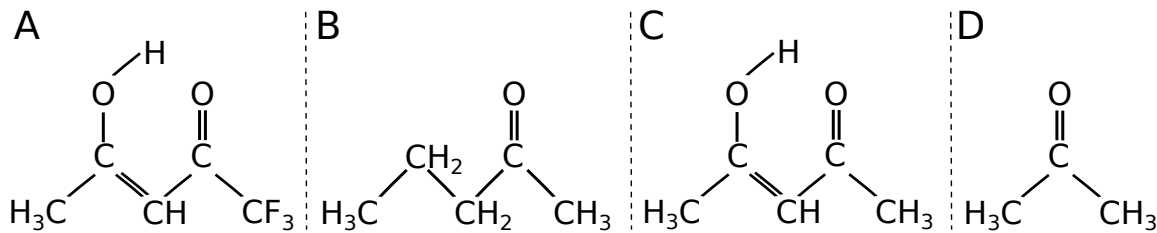}%
\caption{Schematic representations of \textbf{A:} trifluoroacetylacetone (\textit{tfac}), \textbf{B:}  2-pentanone, \textbf{C:} acetylacetone (\textit{acac}), and \textbf{D:} acetone molecules.}
\label{Fig:StructuralFormulas}
\end{figure*}

Set 1 (Fig.~\ref{Fig:Cross-section}A and Fig.~\ref{Fig:Cross-section}C-I) accounts only for the breakage of the metal--ligand bonds of Me$_2$Au(tfac), i.e. for Au--C and Au--O bond breakage due to DI. In this case, the total fragmentation cross section for Me$_2$Au(tfac) has been calculated as a sum of ionization cross sections of its functional groups (see Fig.~\ref{Fig:molecule}) according to the additivity rule principle \cite{DEUTSCH1994}:
\begin{equation}
   \sigma^{\rm Me_2Au(tfac)}_{\rm DI}(E) \approx \sigma^{\rm Au}_{\rm ion}(E) + 2 \sigma^{\rm CH_3}_{\rm ion}(E) + \sigma^{\rm tfac}_{\rm ion}(E) \ .
\label{Eq. TotalIonizationCS}
\end{equation}
Here $\sigma^{\rm Au}_{\rm ion}$, $\sigma^{\rm CH_3}_{\rm ion}$ and $\sigma^{\rm tfac}_{\rm ion}$ are the ionization cross sections for a gold atom \cite{Nelson1976}, CH$_3$ \cite{Hwang1996}, and trifluoroacetylacetone (\textit{tfac}, C$_5$H$_5$F$_3$O$_2$) molecules, respectively. The structure of \textit{tfac} molecule is schematically shown in Fig.~\ref{Fig:StructuralFormulas}A.

The energy-dependent ionization cross section of \textit{tfac}, $\sigma^{\rm tfac}_{\rm ion}(E)$, has been evaluated using the ionization cross section for a structurally similar molecule 2-pentanone (C$_5$H$_{10}$O, see Fig.~\ref{Fig:StructuralFormulas}B)~\cite{Gupta2014}. Assuming a similar shape of the ionization cross sections for \textit{tfac} and 2-pentanone as functions of the projectile kinetic energy $E$, the magnitude of the cross section $\sigma^{\rm tfac}_{\rm ion}(E)$ has been scaled using the ratio of maximum values of cross sections for \textit{tfac} and 2-pentanone:
\begin{eqnarray}
   &&\sigma^{\rm tfac}_{\rm ion}(E) \approx \sigma^{\rm C_5H_{10}O}_{\rm ion}(E) \times \left( \frac{\sigma^{\rm tfac}_{\rm ion, max}}{\sigma^{\rm C_5H_{10}O}_{\rm ion, max}} \right) \\
   &=& \sigma^{\rm C_5H_{10}O}_{\rm ion}(E) \times \left[  \frac{ \left( \sigma_{\rm CH_3}+\sigma_{\rm CF_3}+\sigma_{\rm CH}+2\sigma_{\rm CO} \right)_{\rm max} }{ \left( 2\sigma_{\rm CH_3}+2\sigma_{\rm CH_2}+\sigma_{\rm CO} \right)_{\rm max} } \right]  \nonumber \ .
\label{Eq. tfac_IonizationCS}
\end{eqnarray}
The maximal values of the total ionization cross sections have been evaluated using the functional group and bond additivity model \cite{Bart2001, Bull2012a}. This model is based on a multidimensional matrix least-squares fitting of the correlation between the experimentally measured ionization cross section for 65 organic and halocarbon molecules and the constituent functional groups calculated by means of the Binary-Encounter-Bethe (BEB) model \cite{Hwang1996}. The values of $\sigma^{\rm tfac}_{\rm ion, max}$ and $\sigma^{\rm C_5H_{10}O}_{\rm ion, max}$ have been calculated as a sum of cross section contributions corresponding to CH$_3$, CH$_2$, CH, and CO functional groups.
In Refs.~\cite{Bart2001, Bull2012a}, the maximum values of total electron-impact ionization cross sections for a wide range of halocarbon molecules (including CF$_4$, C$_2$F$_4$, C$_2$F$_6$, C$_3$F$_8$, and others) were evaluated by summing up the partial contributions from C--C and C--F bonds, multiplied by the number of bonds of each type. The calculated cross sections agreed within 10\% accuracy with the corresponding experimental values.
In this study, the contribution from the CF$_3$ functional group to the ionization cross section of \textit{tfac} has therefore been approximated as $\sigma_{\rm CF_3} \approx 3\sigma_{\rm C-F}$, where $\sigma_{\rm C-F}$ is the partial contribution to the ionization cross section from a C--F bond \cite{Bull2012a}.


\subsubsection{Set 2}
\label{sec:Fragm_CS_Set2}

In Set 2 (Fig.~\ref{Fig:Cross-section}B and Fig.~\ref{Fig:Cross-section}C-II), fragmentation of CH$_3$ and \textit{tfac} ligands due to the DEA mechanism has been considered alongside with the breakage of Au--C and Au--O bonds due to DI (considered in Set~1).
The DEA cross section of \textit{tfac} has been evaluated based on the compilation of published data for similar molecules, acetylacetone (\textit{acac}, Fig.~\ref{Fig:StructuralFormulas}C) and acetone (Fig.~\ref{Fig:StructuralFormulas}D).

A comparison of DEA-induced fragmentation mass spectra for \textit{acac} and acetone were reported in Ref.~\cite{Warneke2015a}, showing similar fragmentation patterns for both molecules. In particular, a CHCO$^-$ fragment was detected in the mass spectra for both \textit{acac} and acetone irradiated under the same conditions. Given the absolute cross section for the release of CHCO$^-$ from acetone \cite{Prabhudesai2014}, the ratio of CHCO$^-$ peak intensities in the mass spectra for \textit{acac} and acetone has been used to evaluate the absolute DEA cross section of other fragments from \textit{acac}.

The absolute DEA cross sections for the formation of H$^-$ and CH$_3^-$ fragments from acetone (corresponding to dissociation of C--H and C--CH$_3$ bonds, respectively) have been taken from Ref.~\cite{Prabhudesai2014}.
The absolute DEA cross sections for the formation of F$^-$, CF$_3^-$ and [$\textrm{M} - \textrm{CF}_3$CO]$^-$ fragments (where M denotes the parent Me$_2$Au(tfac) molecule) have been evaluated using the calculated absolute DEA cross section for \textit{acac} and the ratio of relative fragmentation cross sections for \textit{acac} and \textit{tfac} molecules \cite{Omarsson2014}. The partial DEA cross section leading to the breakage of the C--O bond in \textit{tfac} has been evaluated similarly.


\subsubsection{Set 3}
\label{sec:Fragm_CS_Set3}

Set 3 (see Fig.~\ref{Fig:Cross-section}B and Fig.~\ref{Fig:Cross-section}C-III) includes partial DI cross sections of \textit{tfac} and CH$_3$ leading to the cleavage of the C--C, C--F, C--O and C--H bonds. In addition, set 3 includes the partial cross sections of DEA, described above for Set~2.

According to Eq.~(\ref{Eq. DI_CS_approx}), the cross section of DI resulting in the breakage of a C--O bond has been approximated by the ionization cross section of a CO molecule \cite{Hwang1996}.
%
The partial DI cross section leading to the breakage of C--H bonds in CH$_3$ and \textit{tfac} ligands has been calculated using the total ionization cross section of a CH$_3$ molecule \cite{Hwang1996}, divided by three.
The same partial DI cross section has been used for the dissociation of a C--F bond \cite{Bull2012a}.

The study of DI of 2-butanone (CH$_3$COCH$_2$CH$_3$) \cite{Vacher2008} demonstrated that the formation of CH$_3$CO fragments as a result of the ``cental'' C--C bond breakage is the most probable fragmentation channel contributing to 64\% of the total ionization cross section. We have assumed that the \textit{tfac} molecule has a similar fragmentation pattern with the release of CH$_3$CO and CF$_3$CO fragments as the main fragmentation channel. Therefore, the total DI cross section of \textit{tfac} has been split into the contributions leading to the cleavage of the ``central'' C--C bonds ($\sigma_{\rm DI}^{\rm C-CH}$) and the ``side'' C--C bonds ($\sigma_{\rm DI}^{\rm C-CH_3}$ and $\sigma_{\rm DI}^{\rm C-CF_3}$) in approximately the same ratio 2:1 as determined in Ref.~\cite{Vacher2008}. The Me$_2$Au(tfac) molecule contains two ``central'' C--C bonds (C$_3$--C$_4$ bonds in Fig.~\ref{Fig:molecule}) and two ``side''  C--C bonds (C$_1$--C$_4$ and C$_2$--C$_4$ bonds in Fig.~\ref{Fig:molecule}). Therefore, the fragmentation cross section for each bond has been evaluated according to:
\begin{eqnarray}
\sigma_{\rm DI}^{\rm C-CH_3}(E) &\sim& \sigma_{\rm DI}^{\rm C-CF_3}(E) \approx \frac{1}{2} \left( \frac{1}{3} \sigma_{\rm ion}^{\rm tfac}(E) \right) = \frac{1}{6} \sigma_{\rm ion}^{\rm tfac}(E) \ , \nonumber \\
\sigma_{\rm DI}^{\rm C-CH}(E) &\approx& \frac{1}{2} \left( \frac{2}{3} \sigma_{\rm ion}^{\rm tfac}(E) \right) = \frac{1}{3} \sigma_{\rm ion}^{\rm tfac}(E) \ .
\label{Eq. PartialDItfac}
\end{eqnarray}

The partial DI cross sections of Me$_2$Au(tfac) leading to the dissociation of Au--C and Au--O bonds, $\sigma_{\rm DI}^{\rm Au-C}(E)$ and $\sigma_{\rm DI}^{\rm Au-O}(E)$, have been evaluated according to the sum of total ionization cross sections of Au, CH$_3$ and \textit{tfac} fragments, see Eq.~(\ref{Eq. TotalIonizationCS}).
The cross sections $\sigma_{\rm DI}^{\rm Au-C}(E)$ and $\sigma_{\rm DI}^{\rm Au-O}(E)$ have been calculated as follows
\begin{eqnarray}
\sigma_{\rm DI}^{\rm Au-C}(E) &\approx& \frac{1}{3} \sigma_{\rm ion}^{\rm Au} + \sigma_{\rm ion}^{\rm CH_3} \ , \nonumber \\
    \sigma_{\rm DI}^{\rm Au-O}(E) &\approx& \frac{1}{3}\sigma_{\rm ion}^{\rm Au} + \sigma_{\rm ion}^{\rm tfac}
\label{Eq. PartialDIAu}
\end{eqnarray}
to fulfil the sum rule principle for the total DI cross section of Me$_2$Au(tfac), $\sigma^{\rm Me_2Au(tfac)}_{\rm DI}(E) =
2\sigma_{\rm DI}^{\rm Au-C}(E)  + \sigma_{\rm DI}^{\rm Au-O}(E)$, and avoid multiple counting of the contribution $\sigma_{\rm ion}^{\rm Au}$.
The factor 1/3 in Eq.~(\ref{Eq. PartialDIAu}) has been introduced based on the assumption that the contribution of the Au ionization cross section is divided equally between the dissociation channels involving each of two CH$_3$ ligands and the \textit{tfac} ligand.

\subsubsection{Set 4}
\label{sec:Fragm_CS_Set4}

Set 4 uses the same DEA- and DI-induced fragmentation cross sections for all bonds in the Me$_2$Au(tfac) molecule as in Set~3 (see Fig.~\ref{Fig:Cross-section}B and Fig.~\ref{Fig:Cross-section}C-III). Additionally, accounting for chemical reactions involving produced atomic and molecular fragments has been performed in Set~4.
The interactions involving the created fragments can lead to the formation of other volatile molecular species, such as H$_2$, O$_2$, CH$_4$, C$_2$H$_6$ and H$_2$O (see Fig.~\ref{Fig:Cross-section}D). This may affect the number of non-bonded atoms in the deposit and, thus, the resulting metal content.

\subsection{Simulation parameters}

The calculated fragmentation probability $P(x,y)$, Eq.~(\ref{Eq. Frag_Probability_total}), has been tabulated for a $20~{\rm nm} \times 20~{\rm nm}$ grid covering the simulation box and used as input for the IDMD simulations of the irradiation phase of the FEBID process.

Following the earlier studies \cite{Sushko2016, DeVera2020, Prosvetov2021, Prosvetov2021a},
the simulated PE flux density $J_0$ (and hence PE beam current $I_0$) have been rescaled to match the same number of PEs per unit area and per dwell time as in experiments.
This procedure enables the correspondence of simulated results to experimental ones through the correspondence of the electron fluence per dwell time per unit area in simulations and experiments \cite{Sushko2016}.

According to the experimental study of the FEBID of Me$_2$Au(acac) \cite{Mulders2011}, an increase of the electron current $I_{\rm exp}$ from 1.6~nA to 6.3~nA causes minor changes in the elemental composition of the deposits produced by electron irradiation of Me$_2$Au(acac) in the temperature range of $298-423$~K.
Based on those results, the electron current used in the simulations has been set to a characteristic average value $I_{{\rm exp}} =$ 4~nA.
The beam spot radius $R_{{\rm sim}}$ has been set equal to 5~nm and the dwell time value $\tau_d$ has been set to 10~ns, similar to the previous studies \cite{Sushko2016, DeVera2020, Prosvetov2021, Prosvetov2021a}.

The physical state of the system at the end of the replenishment phase of FEBID and prior to the next irradiation phase has been simulated similarly to our earlier IDMD simulations of FEBID \cite{Sushko2016, DeVera2020, Prosvetov2021, Prosvetov2021a}. At first, weakly bound fragments and precursor molecules were removed from the system by an external force field during a 1~ns-long simulation. Afterward, new precursor molecules have been deposited over the circular area with a radius of 7~nm to cover the PE beam spot area and a halo of secondary electrons. Such a model of replenishment prevents the accumulation of non-fragmented molecules along the perimeter of the simulation box where the fragmentation probability is significantly lower \cite{Prosvetov2021a}. The number of precursor molecules added at each FEBID cycle corresponds to the values of the steady-state surface density of Me$_2$Au(tfac) calculated according to Eq.~(\ref{Eq. SteadyStateConcentration}) for each temperature considered in this study.

The simulations have been performed using the Verlet integration algorithm with a time step of 0.5~fs and reflective boundary conditions. Interatomic interactions have been computed using the linked cell algorithm \cite{Solovyov2012, Solo2017} with a cell size of 10~\AA.

\section{Results and discussion}
\label{Results}

\subsection{Analysis of Me$_2$Au(tfac) fragmentation}

The earlier IDMD-based studies of FEBID \cite{DeVera2020, Prosvetov2021} showed that the intensity of precursor fragmentation depends on the number of available fragmentation channels and the amount of energy deposited into the system during the fragmentation process.
A variation of these parameters affects the number and type of molecular fragments produced upon breakage of covalent bonds in the parent molecule.

\begin{figure*}[t!]
	\includegraphics[width=0.8\textwidth]{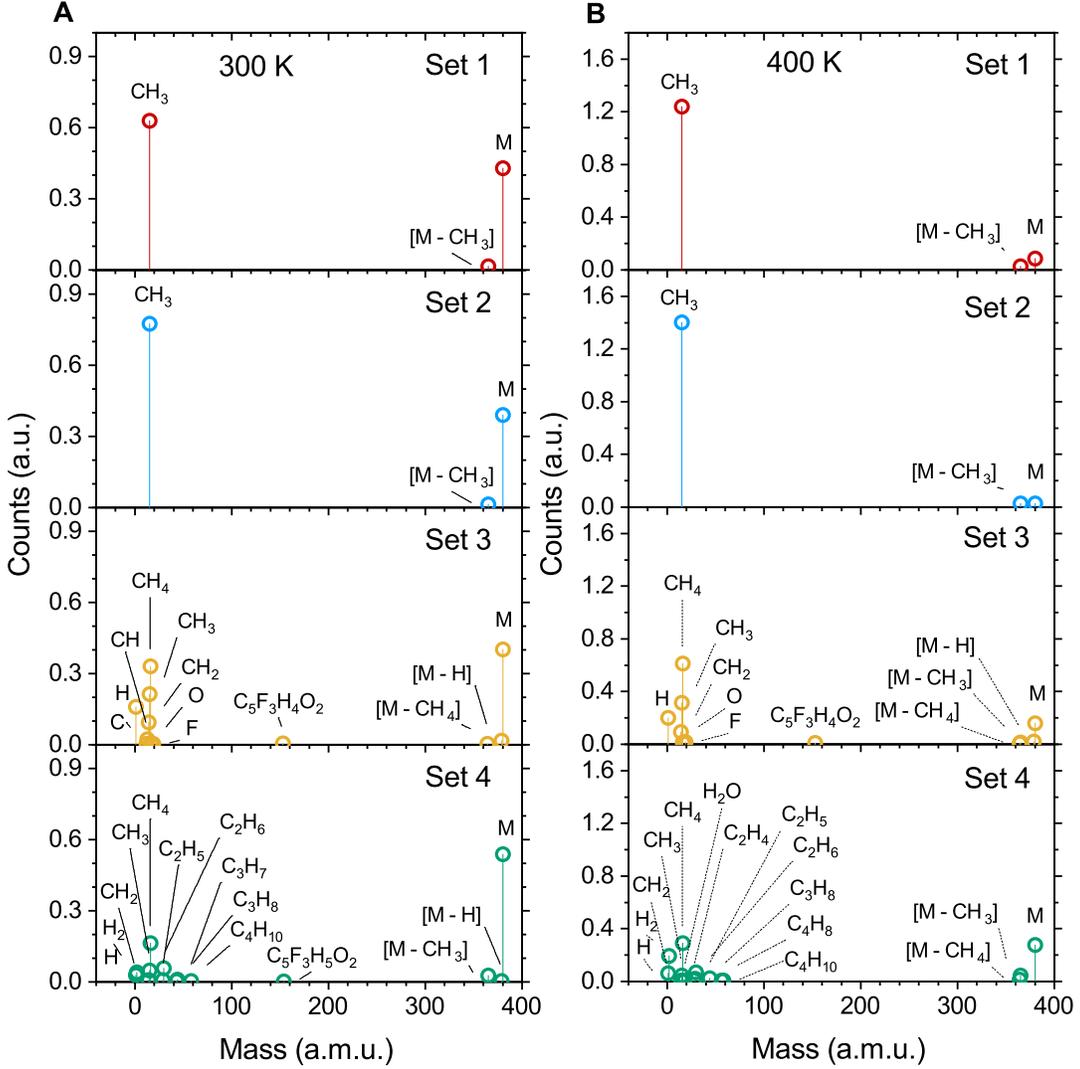}%
	\caption{Relative yields of molecular fragments formed after a 10 ns-long irradiation simulation of adsorbed Me$_2$Au(tfac) molecules (indicated as $M$) at 300~K (panel~\textbf{A}) and 400~K (panel~\textbf{B}) using four sets of fragmentation cross sections described in Sect.~\ref{sec:fragm_CS}. Fragments denoted as $[M - X]$ are produced by the release of a fragment $X$ from the parent molecule $M$. The intensity of each fragment peak is normalized to the number of intact precursor molecules prior to irradiation.}
\label{Fig:MassSpectr}
\end{figure*}

In order to evaluate and analyze the contribution of various DEA and DI fragmentation channels involving different covalent bonds in Me$_2$Au(tfac) to the formation of molecular fragments, four sets of fragmentation cross sections described in Sect.~\ref{sec:fragm_CS} have been considered. Figure~\ref{Fig:MassSpectr} shows the relative yields of Me$_2$Au(tfac) fragments created by the end of a 10-ns long irradiation phase of FEBID. Results of these simulations are summarized also in Fig.~\ref{Fig:Cross-section}D. At a given temperature, the fragment yields shown in Fig.~\ref{Fig:MassSpectr} have been normalized to the number of precursor molecules prior to irradiation for each of the four sets of the fragmentation parameters.

The simulations carried out using Set~1 (top row in Fig.~\ref{Fig:MassSpectr}) indicate the dissociation of Au--C bonds in the Me$_2$Au(tfac) molecule (denoted as M) and the release of CH$_3$ ligands. The addition of fragmentation channels associated with the DEA to CH$_3$ and \textit{tfac} ligands (Set~2, second row) does not lead to any significant change in the relative fragment yield. An explanation for this result is that the corresponding fragmentation cross sections are several orders of magnitude smaller than the partial fragmentation cross sections associated with DI (see Fig.~\ref{Fig:Cross-section}C-II).
In contrast, simulations carried out using the Set~3 (third row in Fig.~\ref{Fig:MassSpectr}) show a larger variety of created fragments. In this case, the dissociation of C--H, C--F and C--O bonds has been observed. The formation of CH$_4$, C$_2$H$_6$ and H$_2$ molecules detected experimentally during electron irradiation of Me$_2$Au(acac) molecules deposited on a surface \cite{Wnuk2010} has been observed in the simulations using only Set~4, in which the formation of C--C, H--H and O--H bonds is enabled by means of the reactive rCHARMM force field. Thus, accounting for DEA and DI fragmentation channels for all the bonds in the Me$_2$Au(tfac) molecule is required to simulate the formation of the experimentally detected molecular fragments. It should be noted that some fragmented molecules have merged in the course of irradiation and formed small clusters containing two or more gold atoms. For the sake of clarity, mass spectra shown in Fig.~\ref{Fig:MassSpectr} are limited by the mass of a parent Me$_2$Au(tfac) molecule, and larger molecular products are not shown.

The results of FEBID simulations carried out at $T = 300$~K (Fig.~\ref{Fig:MassSpectr}A) and 400~K (Fig.~\ref{Fig:MassSpectr}B) demonstrate similar fragmentation patterns for Me$_2$Au(tfac). The fraction of Me$_2$Au(tfac) molecules remaining intact in the entire simulation box after electron-beam irradiation at 400~K is 2 to 5 times lower than at 300~K.
This observation is explained by the difference between the spatial distributions of precursor molecules and the fragmentation rate. Indeed, Me$_2$Au(tfac) molecules are distributed uniformly over the substrate with the surface density depending on the system's temperature. In contrast, the fragmentation rate (independent of temperature) is maximal within the beam spot area with a radius of 5~nm. Towards the edge of the simulation box, the fragmentation rate decreases by several orders of magnitude due to the smaller number of SEs emitted from the substrate in that spatial region. Due to a higher surface density of precursors deposited at 300~K, $n_{\rm p0} = 2.6$~molecules/nm$^2$, most of the molecules outside the beam spot area do not dissociate during one irradiation phase due to the low fragmentation rate in that region. However, a lower surface density of the adsorbed precursors at 400~K ($n_{\rm p0} = 0.8$~molecules/nm$^2$) leads to the fragmentation of almost all the molecules within the beam spot area and in the surrounding halo region by the end of the irradiation phase.

\begin{figure}[t!]
\includegraphics[width=0.5\textwidth]{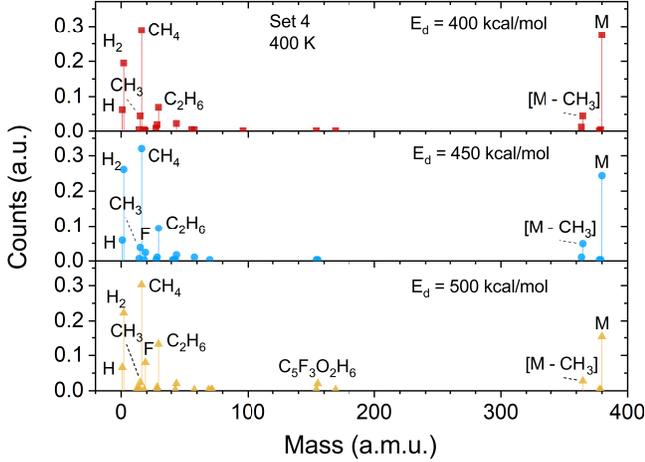}%
\caption{Relative yields of molecular fragments formed after a 10 ns-long irradiation simulation of adsorbed Me$_2$Au(tfac) molecules (denoted as $M$) at $T = 400$~K. Fragments denoted as $[M - X]$ are produced by the release of a fragment $X$ from the parent molecule $M$. The simulations have been conducted using the Set~4 of fragmentation cross sections described in Sect.~\ref{sec:fragm_CS} and the values of the deposited energy $E_d = 400$, 450 and 500~kcal/mol.}
\label{Fig:MassSpectr_Ed}
\end{figure}

Figure~\ref{Fig:MassSpectr_Ed} illustrates the variation of the yield of Me$_2$Au(tfac) fragments for different values of the energy  $E_d$ transferred to the bonded atoms during the fragmentation process. As an illustration, the results are presented for the simulations conducted using Set~4 of fragmentation cross sections described in Sect.~\ref{sec:fragm_CS}. As $E_d$ increases, the fraction of precursor molecules that remain intact by the end of a 10-ns long irradiation phase of FEBID decreases. At the same time, a larger number of fragments (particularly F, C$_2$H$_6$ and C$_5$F$_3$O$_2$H$_6$) is produced.
In general, the value of $E_d$ required for the bond dissociation depends on the molecular structure and environment.
For bulky ligands made of several organic groups (as it happens in $\beta$-diketonates), the energy given to a metal--ligand bond is distributed over many degrees of freedom, thus suppressing the fragmentation process.
As shown below, results of the simulations performed with the values $E_d = 500$~kcal/mol agree with experimental results in terms of the metal content in a deposit.

\subsection{Temperature effects in the FEBID process}

\begin{figure}[t!]
	\includegraphics[width=0.5\textwidth]{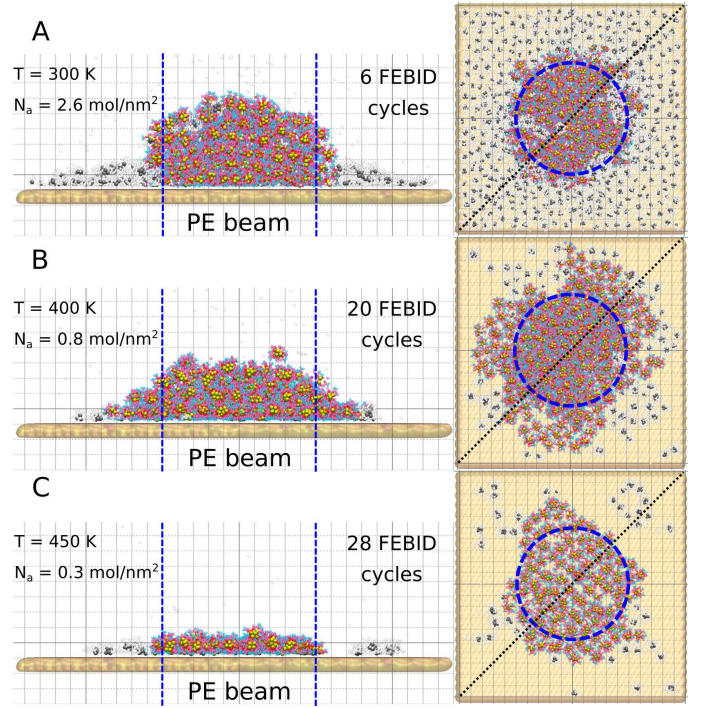}%
	\caption{Snapshots of the IDMD simulations of the FEBID process for Me$_2$Au(tfac) with electron current $I_{\rm exp}=4$~nA at different temperatures $T$ and the corresponding steady-state concentrations of adsorbed precursors $N_{\rm a}$. Left column shows the system's side view on diagonal cross sections indicated by dotted lines on the top view shown in right column.
	The merged largest clusters are visualized in color using the same color scheme as in Fig.~\ref{Fig:molecule}. Isolated precursor molecules and small fragments are shown in gray scale. The primary electron beam spot is depicted by dashed lines in the left column and by circles in the right column. Grid line spacing is 1~nm in all dimensions.}
\label{Fig:Snapshot}
\end{figure}

Temperature at which the FEBID process operates may influence the growth rate and  metal content of deposits \cite{Mulders2011, Koops1996}. A variation in the process temperature also has an impact on the adsorption and diffusion of precursor molecules on a substrate, and their desorption from a substrate. As a result, the equilibrium precursor concentration on the surface depends strongly on temperature.
Figure~\ref{Fig:Snapshot} shows the simulation snapshots of the nanostructures grown at temperatures ranging from 300 to 450~K. As an example, the snapshots are presented for the simulations performed using the fragmentation cross sections from Set~1 (Fig.~\ref{Fig:Cross-section}A). The morphologies of nanostructures obtained by employing other sets of fragmentation cross sections do not show any significant differences to those shown in Fig.~\ref{Fig:Snapshot}. Variation in the number of simulated FEBID cycles at different temperatures is due to the difference in the number of atoms accumulated on a surface.

The largest topologically-connected cluster is shown in Fig.~\ref{Fig:Snapshot} in color, while smaller isolated clusters and intact precursor molecules are presented in gray scale.
Most of precursor molecules adsorbed within the PE beam spot area (indicated in Fig.~\ref{Fig:Snapshot} by dashed lines and circles) undergo fragmentation and merge into a larger structure for all precursor concentrations $n_{\rm p0}$ considered. In comparison with FEBID simulations for Pt(PF$_3$)$_4$ and Fe(CO)$_5$ molecules \cite{Prosvetov2021,Prosvetov2021a}, where the deposited metal clusters merged together forming dendrite-like metal structures, the gold-containing deposit is characterized by small-size metal grains consisting of several gold atoms incorporated into an organic matrix independent of temperature. This difference is explained by the different topology of Pt(PF$_3$)$_4$, Fe(CO)$_5$ and Me$_2$Au(tfac) molecules and the number of non-volatile fragments produced in the course of electron beam irradiation.
The simulation results are in agreement with the experimental analysis of the deposit's morphology for gold-containing $\beta$-diketonate precursors \cite{Riazanova2012, DosSantos2018}. The lateral size of the grown structure depends on the surface density of precursors. At room temperature corresponding to high precursor surface density, the merged structure is limited by the PE beam spot, while it occupies a larger area at $T = 400$~K. The localization of the deposit mostly within the beam spot area at $T = 450$~K (Fig.~\ref{Fig:Snapshot}C) can be explained by a small number of adsorbed precursor molecules.

\begin{figure}[t!]
\includegraphics[width=0.48\textwidth]{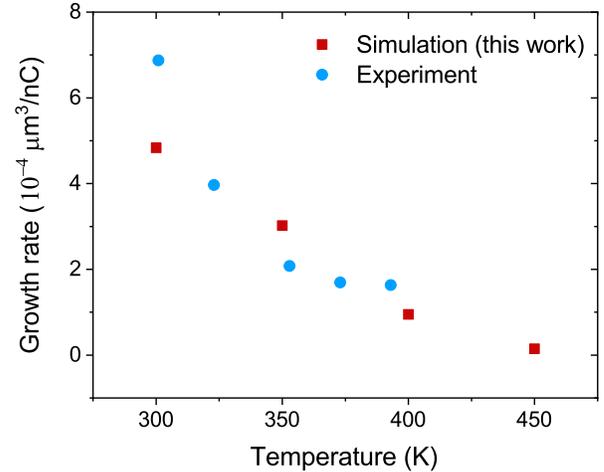}%
\caption{Simulated growth rate of the deposit (red squares) defined as its height in the beam spot area per accumulated electron fluence at different temperatures in the range of $300-450$~K. Blue circles depict the experimental data on the deposit's growth rate during FEBID of Me$_2$Au(acac) \cite{Mulders2011}.}
\label{Fig:Height}
\end{figure}

The deposit's growth rate, defined as the average deposit height per accumulated electron fluence, 
is plotted in Fig.~\ref{Fig:Height} at different temperatures $T$ within the range from 300 to 450~K. The growth rate of the deposit decreases with an increase in the FEBID operating temperature. This result can be explained by a decrease in the steady-state surface density of adsorbed precursors with $T$, see Eq.~(\ref{Eq. SteadyStateConcentration}). Therefore, the number of adsorbed precursor molecules becomes too small to enable the formation of large metal-containing clusters. This simulation result is in agreement with the experimentally measured dependence of the deposit's growth rate on temperature for a structurally similar precursor molecule Me$_2$Au(acac) \cite{Mulders2011}.

\begin{figure}[t!]
\includegraphics[width=0.48\textwidth]{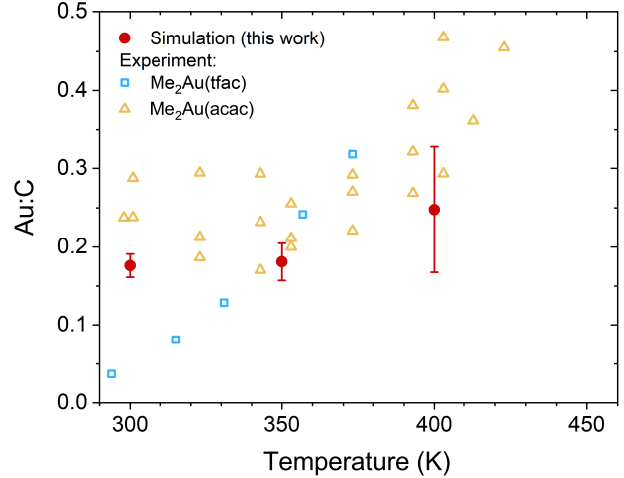}%
\caption{Comparison of the Au:C ratio in the FEBID-grown deposits as a function of treatment temperature. Full symbols correspond to the simulation results for Me$_2$Au(tfac) performed using Set 4 of the fragmentation cross section. Open symbols indicate the experimentally measured Au:C ratios in the deposits grown with Me$_2$Au(tfac) \cite{Koops1996} and Me$_2$Au(acac) \cite{Mulders2011} precursor molecules at different substrate temperatures.}
\label{Fig: Au:C ratio}
\end{figure}

The metal content in deposits is evaluated differently in different FEBID experiments. Some experimental studies reported the Au:C ratio \cite{Koops1996}, while other studies reported the relative fraction of Au, C and O atoms in the deposits \cite{Mulders2011a, DosSantos2018,Shawrav2016a}. In this study, the metal content in the deposit is characterized by the Au:C ratio for easier comparison with the experimental results.
Figure~\ref{Fig: Au:C ratio} compares the results of IDMD simulations (full circles) with experimental data on the Au:C ratio in the deposits of Me$_2$Au(tfac) \cite{Koops1996} and Me$_2$Au(acac) \cite{Mulders2011} (open symbols) as a function of the FEBID process temperature. The Au:C ratio obtained in the simulations is within the range of experimentally reported values and follows the experimentally observed trend that the concentration of gold in the deposit increases with temperature.

Higher metal content in the deposit grown at 400~K can be explained by a combination of several factors. First, the deposition process at elevated temperatures leads to faster thermal desorption of intact precursor molecules and created volatile fragments. As a result, the electron-induced dissociation process takes place in a less dense environment, which leads to a more efficient release of fragments from the deposit. Second, the rates of chemical reactions involving atomic and molecular fragments of Me$_2$Au(tfac) should increase with increasing the temperature at which the FEBID process operates. Third, the operational temperature of FEBID governs the diffusion of precursor molecules and fragments during irradiation. This affects follow-up chemistry and, thus, atomic content, morphology and the growth rate of the deposits.

\section{Conclusions}
\label{Conclusions}

Irradiation-driven molecular dynamics (IDMD) simulations have been performed to explore the role of thermal effects during the FEBID process of Me$_2$Au(tfac), a commonly used precursor molecule for the fabrication of gold nanostructures.

The absolute cross section of electron-induced fragmentation of Me$_2$Au(tfac) required as an input for IDMD has been obtained from the experimentally measured fragmentation mass spectra and fragment ion yields for structurally similar molecules and smaller functional groups of Me$_2$Au(tfac).
The cross section has been evaluated by different methods accounting for DI- and DEA-induced cleavage of different covalent bonds in the molecule. In the simplest approximation, the calculated total fragmentation cross section accounted only for the DI-induced cleavage of covalent bonds between the gold atom and the ligands. The most complete approximation for the fragmentation cross section accounted for the contribution of DI and DEA processes in the cleavage of covalent bonds between the gold atom and the ligands, as well as for the bond cleavage within the ligands.
The explicit simulation of chemical reactions involving the created atomic and molecular fragments has enabled the formation of volatile molecular products H$_2$, CH$_4$ and C$_2$H$_6$ which were observed experimentally during FEBID of Me$_2$Au(acac).

The FEBID process of Me$_2$Au(tfac) precursor molecules has been simulated at different temperatures in the range $300-450$~K.
The simulations confirm experimental observations that deposits consist of small gold clusters embedded into a carbon-rich organic matrix. The simulated growth rate of the deposit decreases from $5 \times 10^{-4}$ to $0.1 \times 10^{-4}$ $\mu$m$^3$/nC upon the temperature increase from 300 to 450~K. A larger number of Me$_2$Au(tfac) fragments created while accounting for the DEA- and DI-induced cleavage of all the bonds in the precursor molecule leads to an increase in the concentration of gold in the deposit. The simulations predict an increase in Au:C ratio in the deposits from $\sim$0.18 to $\sim$0.25 upon increasing the temperature from 300 to 450~K. The simulated deposit's characteristics, such as the deposit's structure, morphology, growth rate, and elemental composition at different temperatures, are in agreement with experimental data.


\begin{acknowledgments}
The authors acknowledge financial support from the Deutsche Forschungsgemeinschaft (Project no. 415716638), and the European Union's Horizon 2020 research and innovation programme – the RADON project (GA 872494) within the H2020-MSCA-RISE-2019 call.
This article is also based upon work from the COST Action CA20129 MultIChem, supported by COST (European Cooperation in Science and Technology).
The possibility of performing computer simulations at the Goethe-HLR cluster of the Frankfurt Center for Scientific Computing is gratefully acknowledged.
\end{acknowledgments}

\bibliography{MBN-RC}

\end{document}